\documentclass[english,aps,prstper,reprint,showpacs,titlepage,longbibliography]{revtex4-2}   

\usepackage[T1]{fontenc}	
\usepackage[latin9]{inputenc}	
\usepackage{geometry}		
\usepackage{graphicx}
\usepackage[above,below]{placeins}	
\usepackage{times}
\usepackage{dirtytalk}
\usepackage{babel}
\usepackage{tabularx}
\usepackage{wrapfig}
\usepackage{orcidlink}
\usepackage{hyperref}  
\hypersetup{colorlinks=true,urlcolor=blue,citecolor=blue,linkcolor=blue}   
\usepackage{enumitem}          
\usepackage{titlesec} 
\setlist{nosep}                 

\begin{document}

\begin{titlepage}
  \title{Power, Prescription, and Postpositivism: Considerations for collecting and representing neurodiversity demographic information in physics education research}
  \author{Mason D. Moenter \orcidlink{0009-0004-0177-2210}}
  \affiliation{Department of Physics \& Astronomy, Texas A\&M University, 578 University Drive College Station, TX 77843}
  \affiliation{Department of Physics \& Astronomy, Michigan State University, 567 Wilson Rd, East Lansing, MI 48824}
  \author{George R. Keefe}
  \affiliation{Physics and Astronomy Department, Rutgers University, 136 Frelinghuysen Rd, Piscataway, NJ 08854}
  \author{Liam G. McDermott \orcidlink{0000-0002-0594-0486}}
  \author{Erin M. Scanlon \orcidlink{0000-0001-8672-7187}} 
  \affiliation{Department of Physics, University of Connecticut $-$ Avery Point, 1084 Shennecossett Rd, Groton, CT, 06340}
  

\begin{abstract}

Demographic data collection is essential in education research, as demographic data allows researchers to better describe the participant population they study and to contextualize findings. However, current research practices for neurodiversity demographics often rely on prescriptive methods (e.g., requiring participants to report official diagnoses) rather than allowing participants to self-identify. This approach can: a) not allow participants to express their intersecting identities in ways that are authentic; and b) limit trustworthiness and reliability of the data and interpretation. In addition, inconsistent dissemination and representation of demographic data across studies hinder the accessibility and usability of this work. Through a literature review of neurodivergent student experiences with learning and performing STEM, we identified widespread discrepancies in how demographic information is collected and reported. This paper explores how neurodivergent identities can be more accurately and inclusively represented in education research. We present findings of a thematic analysis on the ways neurodivergent demographic data collection is done in the literature using data from a systematic literature review on neurodivergent science, technology, engineering, and mathematics (STEM) learning and performance. We call on the PER community to contribute to the development of a framework that centers participant autonomy while supporting clarity, consistency, and future research use.

\clearpage

\end{abstract}
  \maketitle
  
\end{titlepage}

\section{Introduction}


"Who do we study when we study Physics Education Research? (PER)" That is the question asked by Kanim and Cid in their paper \cite{KanimandCid} on the demographic makeup of studies in physics education research (e.g., who do physics education researchers sample from when they study physics learning?). They found that PER overwhelmingly samples from white and wealthy populations, calculus-based courses, and four-year colleges. They also do not mention disability, how ability is represented in PER, or whether or not we even collect data about disabled or neurodivergent students \cite{Frazer2017}. 
Scanlon and Chini \cite{ChiniandScanlon} conducted a literature review of "the nexus of physics, teaching, and disability" (pp 10) and found 16 sources that did not dis-aggregate impairment types, 8 that focused on multiple impairment categories, and 48 sources focused on specific impairment types. To date, there is no published work which examines the ways neurodivergent people are represented in PER. In an attempt to partially remedy this dearth of research, the authors conducted a literature review of neurodivergent STEM learning and performance with the research question of \textit{how are neurodivergent students represented in STEM learning/performance literature?} In doing so, the authors found inconsistencies with how we collect and represent demographic data in education research, especially demographic data specific to neurodivergent populations \cite{PCAST}.


Taking both accurate and usable (able to be used by researchers/practitioners to analyze statistics without loss of or gaps in information \cite{Collins2016, Museus2014}) demographic data is an ethical obligation \cite{Call2022} and a methodological necessity \cite{Brookings2020}. In our literature review, we found that often, researchers took a \textit{prescriptive} rather than a \textit{descriptive} method of collecting and representing disability demographic data. In the following sections, we make the case for why prescriptive demographics are inappropriate for research involving neurodivergent participants, and why descriptive demographics are ethically and methodologically superior. We present findings of our thematic analysis on the demographic data collection of the data corpus of a systematic literature review on neurodivergent STEM learning and performance. Furthermore, we propose a set of guidelines to make demographic data collection/representation more accessible for researchers, inclusive for research participants, and usable for practitioners.

In the following sections we use the words "prescriptivism" and "descriptivism" to describe how neurodivergent demographics are collected and represented in the literature. We specifically borrow these words from linguistics \cite{Pinker2003}, as we found them particularly apt for describing what we found in the literature. We define these terms below:
\begin{itemize}
    \item \textbf{Prescriptivism}- applying preconceived normative categories to data with the implication that the applied categories are the "correct" ones.
    \item \textbf{Descriptivism}- using participants' own words to describe their lived reality.
\end{itemize}


\subsection{Neurodiversity, and Identity}

At the core of any discussion on neurodivergent demographic data is the following question: "what do we mean when we ask 'do you identify as neurodivergent'?" To answer that question, we must also ask: "what does it mean to identify as neurodivergent?" To answer these questions, we turn to the \textit{diversity model of disability} and \textit{post-positivism}.
The diversity model of disability \cite{McDermott2024a, Dwyer2022} describes disability as an interaction (often a fundamental disconnect) between a person's innate ways of thinking, doing, and being and their environment.  When we say someone identifies as neurodivergent, we mean living and experiencing these fundamental disconnects between the environment and oneself.

When we ask "do you identify as neurodivergent?", we unintentionally ask a loaded question. Asking questions in such a manner asserts a categorization (e.g., learning disabled, neurodivergent) and a binary (i.e., someone is either neurodivergent or not) about people's lives. When we collect data in this way, we prescribe strict bounds on participant identity, which may not be reflective of reality. We make assumptions about participant identity which lead to interpretations based on assumptions which are not necessarily reflective of reality, tainting the impacts the findings and dissemination may have. By taking a \textit{postpositivist} \cite{Fox2008,Panhwar2017} approach to analyzing the question "do you identify as neurodivergent?", we must take two key assumptions: 1) participants are actively experiencing a constant re-evaluation and construction of their identity/self based on new ideas, experiences, and situations; and 2) we, as researchers, exist as secondary interpreters of participants' experiences. It is therefore, perhaps, more reflective of reality to ask "\textit{how} do you identify as neurodivergent, if at all?" and build pragmatic categories descriptively, from participants' own words, rather than making assumptions.


\section{Methods}

\subsection{Postionality}

This project, as a part of Author McDermott's postdoctoral fellowship project, is led and conceptualized by Author Moenter. Author Moenter is a neurodivergent, white physics graduate student. Author Keefe is a neurotypical, white physics undergraduate, Author McDermott is a neurodivergent, white and queer physics post-doctoral fellow, and Author Scanlon is a white woman with migraines, depression, and anxiety, who is a physics professor and physics education researcher. They are currently conducting a literature review investigating what extant literature says regarding neurodivergent STEM learning and performance. 

\subsection{Data Collection and Analysis}
This literature review involved Moenter, Keefe, Scanlon, and McDermott sifting through 2,469 sources (peer-reviewed journal articles, conference proceedings, Master's theses, and Doctoral Dissertations) to arrive at a corpus of 47 sources on neurodivergent undergraduate STEM learning and performance.

As a part of data collection, Moenter, Keefe, Scanlon, and McDermott compiled an annotated bibliography which collected, among other things, data on studies' sample sizes, non-neurodivergent demographic data included in an article, and the neurodivergent demographic data included in an article. We conducted a thematic analysis of the literature to categorize demographic data collection and representation methods. To do this, Moenter and McDermott met to comb through the corpus for demographic data collection discussions. They collaboratively inductively coded the data. While there are other facets of identity that are important to attend to, the findings present in this paper are representative of trends in neurodivergent undergraduate STEM learning and performance literature. 

\section{Findings}
Fig. 1 shows a breakdown of demographic representation based on neurodivergent identity category. Some papers listed multiple identity categories, and we counted them multiple times. Therefore, the total article count in Fig. 1 is greater that 47. We found a very high percentage of papers (23.4\%) listed "learning disability" as a broad category or "specific learning disability" (10.6\%). This was particularly troubling as those papers did not break down what constituted as a learning disability, hyper-aggregating participants whose experiences may have been different based on different neurodivergent identities. Many of the studies within the literature corpus did not investigate how participants engaged with their disability, identity, learning, and their interactions \cite{Street2012, Morgan2016, Bundock2021, Dishauzi2016, Madden2021, Pfeifer2020, Marino2020, Pfeifer2023, Brown2023, Jordan2014, Jansen2017, Heiman2004, Taylor2020, Cai2016, Pfeifer2021, Gin2021, Heiman2003, Graves2011, Mulhall2024, Burek2022}. Multiple studies only affirmed that participants were learning disabled, and not \textit{how} participants engaged within the broader category of learning disability or even whether they, themselves, would describe themselves as learning disabled. Using hyper-aggregate categories as a variable in a statistical analysis primarily is gap gazing \cite{Museus2011, Harper2010}, and is not particularly useful for descriptive statistics. 

\begin{figure}[t]
    \centering
    \includegraphics[width=1\linewidth]{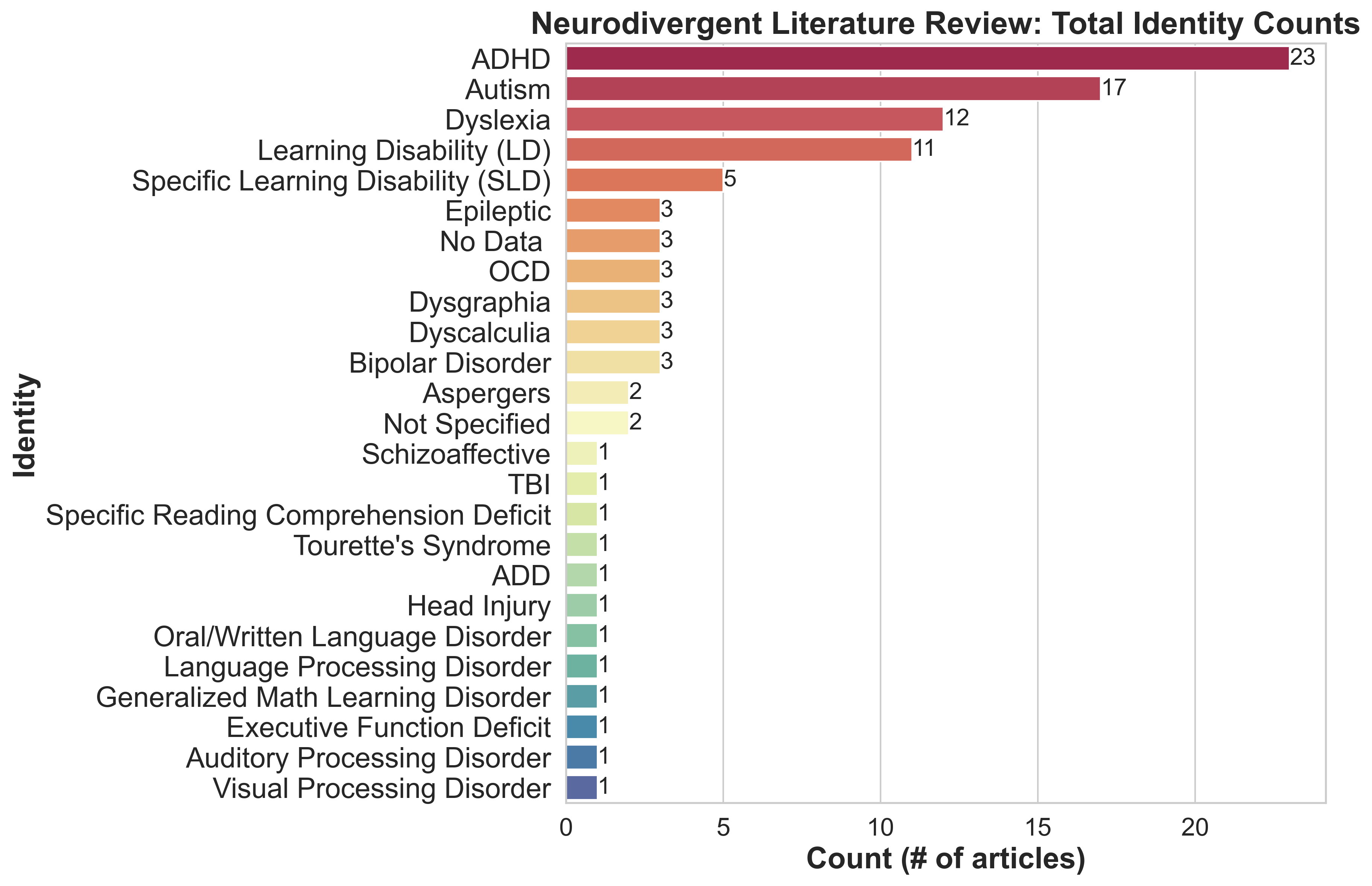}
    \caption{A bar graph of demographic categories included in sources of the literature review. Each count represents a paper in which one or more participants were documented as having a specific identity. Because a single paper can report multiple identities, these counts are not mutually exclusive.}
    \label{fig:1}
\end{figure} 

\subsection{Thematic Analysis Findings}

We found 4 categories of demographic data within the literature corpus.
\begin{enumerate}
    \item \textbf{Descriptive} - demographic categories are built from participants responses.
    \item \textbf{Prescriptive} - demographic categories are pre-determined.
    \item \textbf{Hyper-aggregate} - demographic categories are broad, and often un-detailed.
    \item \textbf{Individual} - demographic categories are small, and often detailed.
\end{enumerate}
We found that the descriptive/prescriptive and the individual/hyper-aggregate categories overlapped in every source, leading us to represent the categories as in Fig. 2. We found that the categories were not a binary, but instead a spectrum; some sources were not purely prescriptive, but still would fall more into the prescriptive end of the descriptive/prescriptive spectrum. The sources which fit in each co-category (e.g. descriptive and hyper-aggregate) are within the individual quadrants in Fig. 2. There are 46 sources included in this categorization because 1 source was a literature review and did not discuss demographic data of their study.

\begin{figure}[t]
    \centering
    \includegraphics[width=1\linewidth]{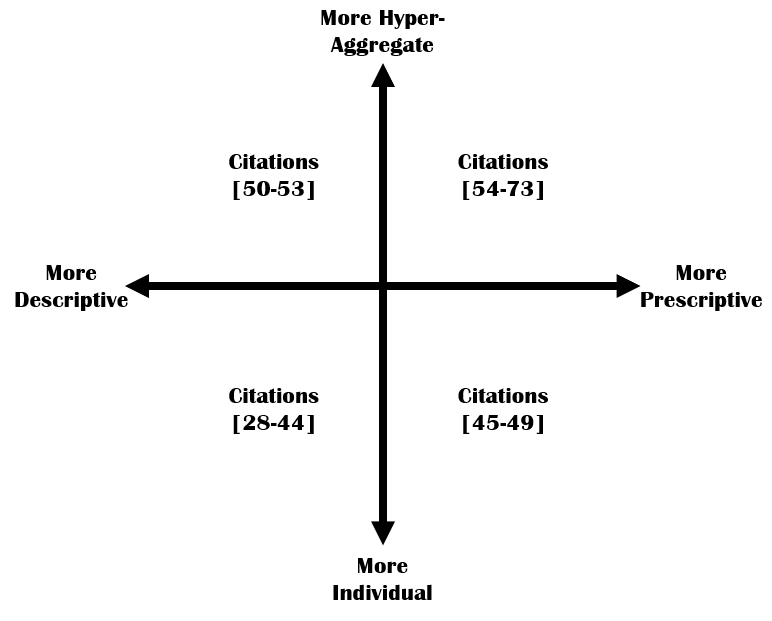}
    \caption{The four demographic data categories visualized in a 2-D chart. The horizontal axis represents the descriptive/prescriptive spectrum of sources, while the vertical axis represents the individual/hyper-aggregate spectrum of sources. The numbers in each of the four quadrant represent the citations which fall into each dual-category.}
    \label{fig:2}
\end{figure}

As can be seen in Fig. 2, a large amount of papers (N=19, 41.3\%) fell into the hyper-aggregate-prescriptive category and (N=16, 34.8\%) fell into the individual-descriptive category. We found that the individual-descriptive sources \cite{Cox2019,Salvatore2024,McDermott2024b,Nieminen2019,Cuellar2022,James2019,James2020,Morris2022,Lapraik2013,Nieminen2024,Mcpeake2023,Zaghi2023,McDermott2023,Salty2022,Wang2024,Gin2022a,Perkin2007} took an approach to demographic data that focused on the lived reality of participants in their own words. We found that the individual-prescriptive sources \cite{Bolourian2018, Accardo2019a, Anderson2018, Accardo2019b, Andreassen2017} took an approach to demographic data that focused on lived experiences within external (not of participants' own words) bounds. We found that the hyper-aggregate-descriptive \cite{Friedensen2021, Taylor2022, Osborne2019, Gin2022b} sources took an approach to demographic data that, while focusing on participants' lived experiences to generate broad categories, provided little to no differentiation between student experiences (focused on a hypothetical amalgamate "the neurodivergent student"). The hyper-aggregate-prescriptive sources \cite{Street2012, Morgan2016, Bundock2021, Dishauzi2016, Madden2021, Pfeifer2020, Marino2020, Pfeifer2023, Brown2023, Jordan2014, Jansen2017, Heiman2004, Taylor2020, Cai2016, Pfeifer2021, Gin2021, Heiman2003, Graves2011, Mulhall2024, Burek2022} took an approach to demographic data that provided little to no differentiation about participant experiences as neurodivergent, and prescribed strict bounds to participant identity.

In a 2024 PERC Proceedings paper, Stella \textit{et al.} \cite{Stella} provide commentary and recommendations for racial demographic data. They call out hyper-aggregation of racial groups into single monolithic categories and state that researchers should treat racial categories as subjective, constructed things, rather than objective truth. We see similar issues reflected in neurodivergent demographic data collection across the literature corpus.

\subsection{Implications}
Experts in both academic and private sectors acknowledge the benefits of collecting demographic information as a part of human research \cite{Brookings2020}. Accurate demographic data collection is essential to ensure that our datasets and studies are truly representative of the populations we study \cite{Sharghi2024}. Demographic data provides the foundation for recognizing and including marginalized communities and underrepresented groups \cite{Museus2011}. Without accurate data, these groups risk being rendered invisible \cite{Brookings2020}. Collecting and analyzing this information allows us to identify disparities, drive meaningful change, and develop more inclusive pedagogical practices. It also enables the physics community to actively challenge ableism and better celebrate the full diversity of experiences and identities that enrich our field. Though just over a third (34.8\%) of sources fell into the descriptive-individual category, meaning they took the effort to represent participants in their own words without hyper-aggregating them, just under two-thirds (65.2\%) of sources hyper-aggregated demographic data, prescribed their own categories of participant identities, or both. Clearly there are inconsistencies in demographic data collection and representation. These inconsistencies can have real, potentially harmful implications for research.

When discussing real human people doing real human things in real human ways, a certain amount of mess is to be expected. Identity is not bound by constrictive categories, and an individual's identity is certainly not bound by a category imposed by a researcher. In prescribing the participant's disability identity, the researcher exerts an undue amount of power which not only serves to marginalize participants - saying that their experiences are not relevant to the question at hand - but to misrepresent reality by calling the participant something which they may not necessarily be. When researchers hyper-aggregate, they wash out unique experiences across people in a group.  This is both ethically contentious and empirically unsound \cite{Mertens2009,Baglieri2010}. The issue, therefore, at hand is to create a standard of collecting and representing disability demographic data in a way which is reflective of individual reality \textit{and} maintains scientific utility for other researchers and educators to consume and iterate upon.

As Stella \textit{et al.} \cite{Stella} describe, demographic categories are not objective truths about the human experience. Demographic categories are just that: categories. The categories that we use must be both descriptive of reality and useful to researchers. As such, categorization, like much of language, is an approximation of reality. We, as researchers need to ensure that we are describing not prescribing the categories. We must build the categories descriptively, not prescriptively.

\section{Building Neurodivergent Identity Data Guidelines (NEURO-ID)}

The prescriptivist approach looks like assigning categories or umbrella terms of LD or neurodivergent, and just leaving those categories for the reader to interpret; with no discussion of how participants are neurodivergent/LD. Practically, this top-down approach involves asking students \textit{if} they identify a certain way, with no further questions. In contrast, the descriptivist lens acknowledges the post-positivist lens on reality as socially constructed. By asking participants how they identify (potentially "how do you identify as neurodivergent?"), researchers allow for the vast diversity of experience to be expressed in participants' own words. Using participants' own descriptions of themselves, researchers can then build categories that are reflective of reality and usable by readers. 

\begin{figure}[htbp]
    \centering
    \hspace{0cm} 
    \includegraphics[width=1\linewidth]{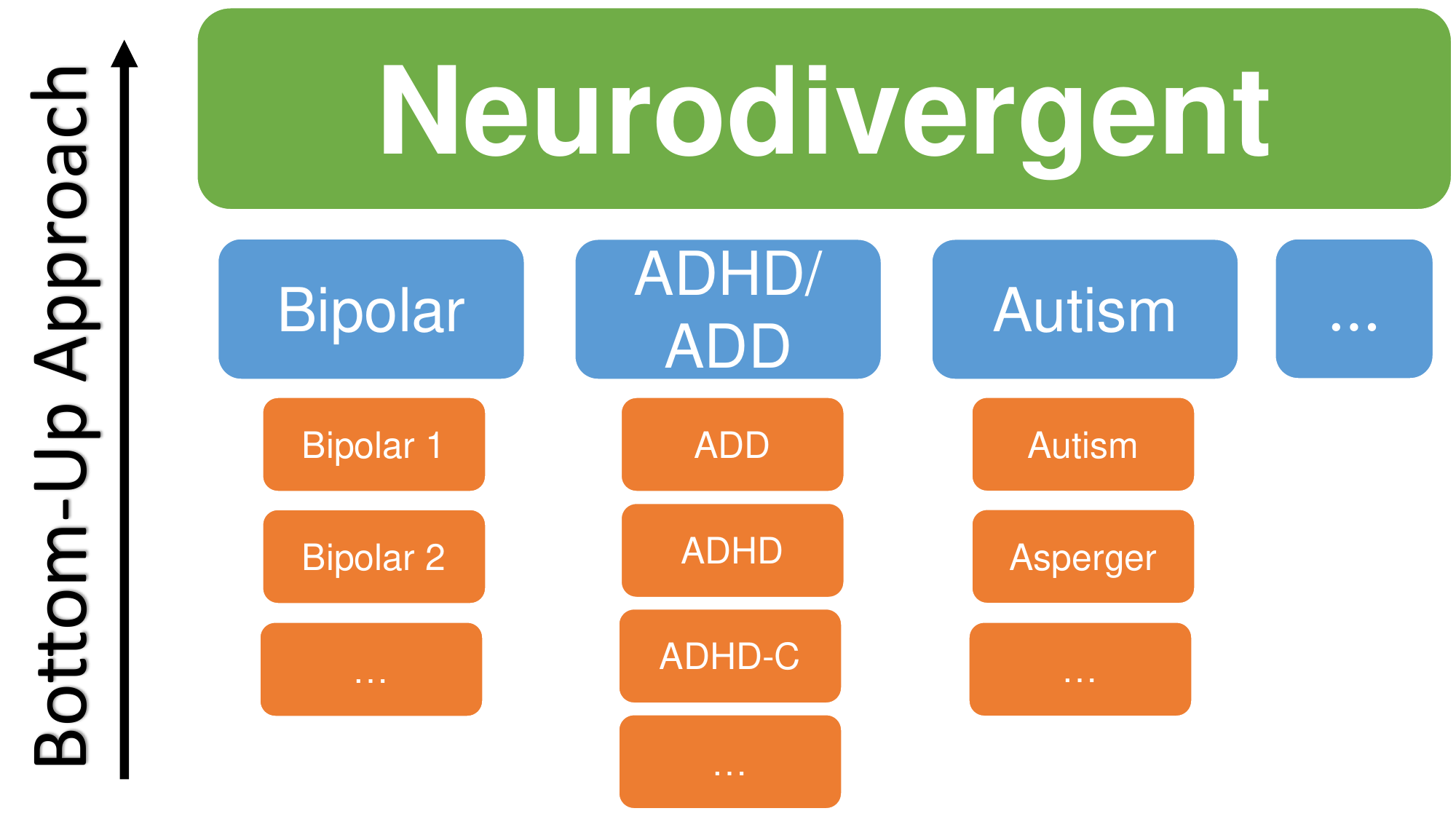}
    \caption{The NEURO-ID approach to demographic data collection. A "bottom-up" (start from the bottom and work your way up) descriptivist approach to demograpic data collection. Scholars start with participant self-identified demographics and, if applicable, generate categories from participant's own descriptions.}
    \label{fig:3}
\end{figure}

The descriptivist approach in Fig. 3 involves forming participants individual identity descriptions into umbrella categories. These umbrella categories, where needed, can be further aggregated into overarching umbrella terms, such as disabled or neurodivergent. To avoid hyper-aggregation, it is critical that when researchers report demographic data in their studies, they report all the categories, not just the topmost category, as that defeats the whole purpose of the descriptivist approach. 

\subsection{Aligning the Guidelines with DisCrit}
To construct the guidelines, we employ the tenets of DisCrit \cite{Annamma2012}. We use DisCrit to inform decisions regarding demographic data collection in ways that are sensitive to neurodivergent participants and are sensitive to the ways disability/neurodivergence intersect with other forms of identity. The following tenets are directly from Annamma \textit{et al.} \cite{Annamma2012}. The recommendations that we create following each tenet are to be used in conjunction with the descriptivist approach described in the previous sections.
\newline
\noindent
\textit{DisCrit focuses on ways that the forces of racism and ableism circulate interdependently, often in neutralized and invisible ways, to uphold notions of normalcy.}
\begin{itemize}
     \item We suggest that scholars take up additional suggestions related to race/ethnicity data collection \cite{Stella}.
\end{itemize}
\textit{DisCrit values multidimensional identities and troubles singular notions of identity such as race or dis/ability or class or gender or sexuality, and so on.}
    \begin{itemize}
        \item We suggest that scholars include disability/neurodivergent identity within their demographic data collection methods for all kinds of studies, not just those focused on disability \cite{ChiniandScanlon}.
    \end{itemize}
\textit{DisCrit emphasizes the social constructions of race and ability and yet recognizes the material and psychological impacts of being labeled as raced or dis/abled, which sets one outside of the western cultural norms.}
    \begin{itemize}
        \item We suggest scholars take a descriptivist approach to demographic data collection, building categories based on participants' own descriptions of their identity
        \item We suggest that scholars take a social-constructivist/post-positivist approach to identity when constructing demographic questions.
        \item We suggest that scholars take up an intracategorical approach to describing and representing demographic data \cite{Baur2019,Vaccaro2020, Goldberg2017, Sommo2013}.
    \end{itemize}
\textit{DisCrit privileges voices of marginalized populations, traditionally not acknowledged within research.}
    \begin{itemize}
        \item We suggest scholars create open-ended demographic questions to encourage participants to use their own words in their description of identity.
        \item We suggest that scholars include neurodivergent scholars on research teams and in the development of study materials/questions, not just as research participants.
        \end{itemize}     
\textit{DisCrit considers legal and historical aspects of dis/ability and race and how both have been used separately and together to deny the rights of some citizens.}
    \begin{itemize}
        \item We suggest scholars consider the ways that demographic categorization has been used to harm disabled and neurodivergent people \cite{Dwyer2022, Powell2020} and specifically use dissemination methods not aligned with medical diagnosis categories to reject the medical model, and use dissemination methods that view disability as diversity. 
        \item We suggest future research specifically at the intersection of disability/neurodivergence and race/ethnicity. This can be done by asking questions about race/ethnicity and disability \textit{and} the intersection of both.
    \end{itemize}
\textit{DisCrit recognizes whiteness and Ability as Property and that gains for people labeled with dis/abilities have largely been made as the result of interest convergence of white, middle-class citizens.}
    \begin{itemize}
        \item We suggest that scholars both weave neurodivergence into studies not otherwise about ability, \textit{and} create studies specifically about disability/neurodivergence. These latter studies would expand data collection beyond disability/ability binaries and likely involve collection of thick descriptions of lived experiences of neurodivergent folks. 
    \end{itemize}
\textit{DisCrit requires activism and supports all forms of resistance.}
    \begin{itemize}
        \item We suggest that researchers take up harm reduction in their research actions (prioritizing and supporting the needs of participants which includes accessible and humane research practices \cite{McPadden2023}); and social-justice orientations to research.
    \end{itemize}


\acknowledgments{
This work was funded by the National Science Foundation (STEM Ed IPRF \#2411711, DUE \#2336367, DUE \#2336368), the American Physical Society Forum on Education (FEd) Project Mini-Grant, and the Departments of Physics \& Astronomy at the University of Connecticut and Michigan State University. 

Special thanks to the Courses 2 Careers design team and the BECOLA group at FRIB.

We also thank the team of Houseplant and Lovelace McDermott-Moore for their support in writing this article.
}




\newpage
\begin{thebibliography}{99}
  \bibitem{KanimandCid}Kanim, S., \& Cid, X. C. (2020). Demographics of physics education research. \textit{Physical Review Physics Education Research, 16(2), 020106.} 
  
  \bibitem{Frazer2017}Frazer, L. (2017). Diversity in physics education research: A decade of inclusion and future challenges. \textit{arXiv preprint arXiv:1710.07863.} 
  
  \bibitem{ChiniandScanlon}Chini, J. J., \& Scanlon, E. M. (2023). Teaching physics with disabled learners: A review of the literature. \textit{The International Handbook of Physics Education Research: Special Topics, 1-1.} 
  
  \bibitem{PCAST}President's Council of Advisors on Science and Technology. (2024). Harnessing the power of social and behavioral science to improve American lives. \textit{Executive Office of the President. https://www.whitehouse.gov/pcast}

  \bibitem{Collins2016}Collins, P. H., \& Bilge, S. (2016). \textit{Intersectionality. Polity Press.}

  \bibitem{Museus2014}Museus, S.D. (2014). The Culturally Engaging Campus Environments (CECE) Model: A New Theory of Success Among Racially Diverse College Student Populations. \textit{In: Paulsen, M. (eds) Higher Education: Handbook of Theory and Research. Higher Education: Handbook of Theory and Research, vol 29. Springer, Dordrecht. https://doi.org/10.1007/978-94-017-8005-6}

  \bibitem{Call2022}Call, C. C., Eckstrand, K. L., Kasparek, S. W., Boness, C. L., Blatt, L., Jamal-Orozco, N., Novacek, D. M., \& Foti, D. (2022). An Ethics and Social-Justice Approach to Collecting and Using Demographic Data for Psychological Researchers. \textit{Perspectives on Psychological Science, 18(5), 979-995. https://doi.org/10.1177/17456916221137350 (Original work published 2023)}
  
  \bibitem{Brookings2020}R. Ray, (2020). The importance of collecting demographic data. \textit{The Brookings Institute}.

  \bibitem{Pinker2003}Pinker, S. (2003). The language instinct: How the mind creates language. \textit{Penguin uK.}
  
  \bibitem{McDermott2024a}Liam Gregory McDermott; Introducing Neurodiversity to the Physics Education Community. \textit{Phys. Teach. 1 September 2024; 62 (6): 472-475. https://doi.org/10.1119/5.0135030}

  \bibitem{Dwyer2022}Dwyer, P. (2022). The neurodiversity approach (es): What are they and what do they mean for researchers?. \textit{Human development, 66(2), 73-92. https://doi.org/10.1159/000523723}

  \bibitem{Shakespeare2006}Shakespeare, T. (2006). The social model of disability. In The disability studies reader (pp. 16-24). Routledge.
  
  \bibitem{Fox2008}N. J. Fox, (2008). Post-positivism. \textit{The SAGE encyclopedia of qualitative research methods, 2(1), 659-664.} 
  
  \bibitem{Panhwar2017}Panhwar, A. H., Ansari, S., \& Shah, A. A. (2017). Post-positivism: An effective paradigm for social and educational research. \textit{International Research Journal of Arts and Humanities, 45(45), 253-259.} 

  \bibitem{Sharghi2024}Sharghi, S., Khalatbari, S., Laird, A., Lapidus, J., Enders, F. T., Meinzen-Derr, J., Tapia, A. L., \& Ciolino, J. D. (2024). Race, ethnicity, and considerations for data collection and analysis in research studies. \textit{Journal of clinical and translational science, 8(1), e182. https://doi.org/10.1017/cts.2024.632}

  \bibitem{Museus2011}Museus, S.D. and Griffin, K.A. (2011), Mapping the margins in higher education: On the promise of intersectionality frameworks in research and discourse. \textit{New Directions for Institutional Research, 2011: 5-13. https://doi.org/10.1002/ir.395}

  \bibitem{Harper2010}Harper, S. R. (2010). An anti-deficit achievement framework for research on students of color in STEM. \textit{New directions for institutional research, 2010(148), 63-74.} 
  
  \bibitem{Mertens2009}Mertens, D. M. (2009). \textit{Transformative research and evaluation. New York: Guilford Press.}

  \bibitem{Baglieri2010}Baglieri, S., Valle, J. W., Connor, D. J., \& Gallagher, D. J. (2010). Disability Studies in Education: The Need for a Plurality of Perspectives on Disability. \textit{Remedial and Special Education, 32(4), 267-278. https://doi.org/10.1177/0741932510362200 (Original work published 2011)} 

  \bibitem{Stella}Stella, S., Robertson, A., \& V\'elez, V. (2024, July 10-11). Considerations for Collecting Racial Demographics Data in Physics Education Research. \textit{Paper presented at Physics Education Research Conference 2024, Boston, MA. Retrieved May 19, 2025, from https://www.compadre.org/Repository/document/ServeFile.cfm, ID=16929\&DocID=5996}

  \bibitem{Annamma2012}Annamma, S. A., Connor, D., \& Ferri, B. (2012). Dis/ability critical race studies (DisCrit): theorizing at the intersections of race and dis/ability. \textit{Race Ethnicity and Education, 16(1), 1-31. https://doi.org/10.1080/13613324.2012.730511}

  \bibitem{Baur2019}Greta R. Bauer, Ayden I. Scheim, Advancing quantitative intersectionality research methods: Intracategorical and intercategorical approaches to shared and differential constructs, \textit{Social Science \& Medicine, Volume 226, 2019, Pages 260-262, ISSN 0277-9536, https://doi.org/10.1016/j.socscimed.2019.03.018.}

  \bibitem{Vaccaro2020}Vaccaro, A., Lee, M. N., Tissi-Gassoway, N., Kimball, E. W., \& Newman, B. M. (2020). Gender and Ability Oppressions Shaping the Lives of College Students: An Intracategorical, Intersectional Analysis. \textit{Journal of Women and Gender in Higher Education, 13(2), 119-137. https://doi.org/10.1080/26379112.2020.1780134}

  \bibitem{Goldberg2017}Goldberg, C. (2016). Is Intersectionality a Disabled Framework? Presenting PWIVID: In/Visibility and Variability as Intracategorical Interventions. \textit{Critical Disability Discourses, 7. https://doi.org/10.25071/1918-6215.39695}

  \bibitem{Sommo2013}Sommo, A. and Chaskes, J. (2013), "Intersectionality and the disability: Some conceptual and methodological challenges", \textit{Disability and Intersecting Statuses (Research in Social Science and Disability, Vol. 7), Emerald Group Publishing Limited, Leeds, pp. 47-59. https://doi.org/10.1108/S1479-3547(2013)0000007005}

  \bibitem{Powell2020}Powell, R. M. (2020). Confronting eugenics means finally confronting its ableist roots. \textit{Wm. \& Mary J. Race Gender \& Soc. Just., 27, 607.https://ssrn.com/abstract=3706408}

  \bibitem{McPadden2023}McPadden, D., Sawtelle, V., Scanlon, E. M., Chini, J. J., Chahal, H., Levy, R., \& Reynolds, A. (2023). Planning for participants' varying needs and abilities in qualitative research. \textit{Physical Review Physics Education Research, 19(2), 020143.} 

  \bibitem{Cox2019}Cox, Thomas \& Ogle, Brian \& Campbell, Laurie. (2019). Investigating Challenges and Preferred Instructional Strategies in STEM. \textit{Journal of Postsecondary Education and Disability, 32(1) 49-61} 

  \bibitem{Salvatore2024}Salvatore, S., White, C. \& Podowitz-Thomas, S. (2024). Not a cookie cutter situation: how neurodivergent students experience group work in their STEM courses. \textit{IJ STEM Ed 11, 47 https://doi.org/10.1186/s40594-024-00508-0} 

  \bibitem{McDermott2024b}McDermott, L. G., Mosley, N. A., \& Cochran, G. L. (2024). Diverging nonlocal fields: Operationalizing critical disability physics identity with neurodivergent physicists outside academia. \textit{Physical Review Physics Education Research, 20(1), 010111. https://doi.org/10.1103/PhysRevPhysEducRes.20.010111}  
   
  \bibitem{Nieminen2019}Nieminen, J. H., \& Pesonen, H. V. (2019). Taking universal design back to its roots: Perspectives on accessibility and identity in undergraduate mathematics. \textit{Education Sciences,.10(1), 12. https://doi.org/10.3390/educsci10010012} 

  \bibitem{Cuellar2022}Cuellar, A., Webster, B., Solanki, S., Spence, C., \& Tsugawa, M. (2022, August). Examination of ableist educational systems and structures that limit access to engineering education through narratives. \textit{In 2022 ASEE Annual Conference \& Exposition. https://peer.asee.org/41800}

  \bibitem{James2019}James, W., Lamons, K., Spilka, R., Bustamante, C., Scanlon, E., \& Chini, J. J. (2019). Hidden walls: STEM course barriers identified by students with disabilities. \textit{arXiv preprint: https://arxiv.org/abs/1909.02905}

  \bibitem{James2020}James, W., Bustamante, C., Lamons, K., Scanlon, E., \& Chini, J. J. (2020). Disabling barriers experienced by students with disabilities in postsecondary introductory physics. \textit{Physical Review Physics Education Research, 16(2), 020111. https://doi.org/10.1103/PhysRevPhysEducRes.16.020111}

  \bibitem{Morris2022}Morris, N. (2022). "This Isn't Working for Me. Can We Do It a Different Way?" The Lived Experiences of Geoscience Students with Learning Disabilities. \text{Dissertation: Western Michigan University.}

  \bibitem{Lapraik2013}Lapraik, S. (2013). Dyslexic students preparing for examinations in higher education: strategies and a sense of control. \textit{Doctoral dissertation, University of Southampton}

  \bibitem{Nieminen2024}Nieminen, J. H., Reinholz, D. L., \& Valero, P. (2024). "Mathematics is a battle, but I've learned to survive": becoming a disabled student in university mathematics. \textit{Educational Studies in Mathematics, 116(1), 5-25. https://doi.org/10.1007/s10649-024-10311-x}

  \bibitem{Mcpeake2023}Mcpeake, E., Lamore, K., Boujut, E., El Khoury, J., Pellenq, C., Plumet, M. H., \& Cappe, E. (2023). "I just need a little more support": A thematic analysis of autistic students' experience of university in France. \textit{Research in Autism Spectrum Disorders, 105, 102172. https://doi.org/10.1016/j.rasd.2023.102172}

  \bibitem{Zaghi2023}Zaghi, A. E., Grey, A., Hain, A., \& Syharat, C. M. (2023). "It Seems Like I'm Doing Something More Important"-An Interpretative Phenomenological Analysis of the Transformative Impact of Research Experiences for STEM Students with ADHD. \textit{Education Sciences, 13(8), 776. https://doi.org/10.3390/educsci13080776}

  \bibitem{McDermott2023}McDermott, L. G., \& Mosley, N. A. (2023, September). "Academia, as a whole, is structured entirely without any consideration for neurodivergency," and other things neurodivergent students want you to know. \textit{In Proc. Phys. Educ. Res. Conf. https://doi.org/10.1119/perc.2023.pr.McDermott}

  \bibitem{Salty2022}Salty, K., Gobernatz, A., \& Close, E. (2022, September). ADH... Disorder? Discoveries on ADHD and physics learning from collaborative autoethnography. \textit{In Proc. Phys. Educ. Res. Conf. doi.org/10.1119/perc.2022.pr.Salty}

  \bibitem{Wang2024}Wang, K. D., McCool, J., \& Wieman, C. (2024). Exploring the learning experiences of neurodivergent college students in STEM courses. \textit{Journal of Research in Special Educational Needs, 24(3), 505-518. https://doi.org/10.1111/1471-3802.12650}

  \bibitem{Gin2022a}Gin, L. E., Pais, D., Cooper, K. M., \& Brownell, S. E. (2022). Students with disabilities in life science undergraduate research experiences: Challenges and opportunities. \textit{CBE-Life Sciences Education, 21(2), ar32. https://doi.org/10.1187/cbe.21-07-0196}

  \bibitem{Perkin2007}Perkin, G., \& Croft, T. (2007). The dyslexic student and mathematics in higher education. \textit{Dyslexia, 13(3), 193-210. https://doi.org/10.1002/dys.334}

  \bibitem{Bolourian2018}Bolourian, Y., Zeedyk, S. M., \& Blacher, J. (2018). Autism and the university experience: Narratives from students with neurodevelopmental disorders.\textit{Journal of autism and developmental disorders, 48, 3330-3343. https://doi.org/10.1007/s10803-018-3599-5}

  \bibitem{Accardo2019a}Accardo, A. L., Kuder, S. J., \& Woodruff, J. (2019). Accommodations and support services preferred by college students with autism spectrum disorder.\textit{Autism, 23(3), 574-583. https://doi.org/10.1177/1362361318760490} 

  \bibitem{Anderson2018}Anderson, A. H., Carter, M., \& Stephenson, J. (2018). Perspectives of university students with autism spectrum disorder. \textit{Journal of autism and developmental disorders, 48, 651-665. https://doi.org/10.1007/s10803-017-3257-3} 

  \bibitem{Accardo2019b}Accardo, A. L., Bean, K., Cook, B., Gillies, A., Edgington, R., Kuder, S. J., \& Bomgardner, E. M. (2019). College access, success and equity for students on the autism spectrum. \textit{Journal of autism and developmental disorders, 49, 4877-4890. https://doi.org/10.1007/s10803-019-04205-8}

  \bibitem{Andreassen2017}Andreassen, R., Jensen, M. S., \& Br\aa ten, I. (2017). Investigating self-regulated study strategies among postsecondary students with and without dyslexia: A diary method study. \textit{Reading and Writing, 30, 1891-1916. https://doi.org/10.1007/s11145-017-9758-9}

  \bibitem{Friedensen2021}Friedensen, R., Lauterbach, A., Kimball, E., \& Mwangi, C. G. (2021). Students with high-incidence disabilities in STEM: Barriers encountered in postsecondary learning environments. \textit{Journal of Postsecondary Education and Disability, 34(1), 77-90. https://eric.ed.gov/?id=EJ1308649}

  \bibitem{Taylor2022}Taylor, C. L., \& Zaghi, A. E. (2022). The interplay of ADHD characteristics and executive functioning with the GPA and divergent thinking of engineering students: A conceptual replication and extension. \textit{Frontiers in Psychology, 13, 937153. https://doi.org/10.3389/fpsyg.2022.937153}

  \bibitem{Osborne2019}Osborne, T. (2019). Not lazy, not faking: teaching and learning experiences of university students with disabilities. \textit{Disability \& Society, 34(2), 228-252. https://doi.org/10.1080/09687599.2018.1515724}

  \bibitem{Gin2022b}Gin, L. E., Pais, D. C., Parrish, K. D., Brownell, S. E., \& Cooper, K. M. (2022). New online accommodations are not enough: The mismatch between student needs and supports given for students with disabilities during the COVID-19 pandemic. \textit{Journal of microbiology \& biology education, 23(1), e00280-21. https://doi.org/10.1128/jmbe.00280-21} 
  
  \bibitem{Street2012}Street, C. D., Koff, R., Fields, H., Kuehne, L., Handlin, L., Getty, M., \& Parker, D. R. (2012). Expanding Access to STEM for At-Risk Learners: A New Application of Universal Design for Instruction. \textit{Journal of Postsecondary Education and Disability, 25(4), 363-375. https://eric.ed.gov/?id=EJ1002146}

  \bibitem{Morgan2016}Morgan, M. V. C. (2016). The STEM and CTE Pipeline for Community College Students with Learning Disabilities. \textit{Dissertation: University of California, Santa Barbara.}

  \bibitem{Bundock2021}Bundock, K., Callan, G. L., Longhurst, D., Rolf, K. R., Benney, C. M., \& McClain, M. B. (2021). Mathematics Intervention for College Students With Learning Disabilities: A Pilot Study Targeting Rate of Change. \textit{Insights into Learning Disabilities, 18(1), 1-28. https://eric.ed.gov/?id=EJ1295246} 

  \bibitem{Dishauzi2016}Dishauzi, K. M. (2016). Supporting students with disabilities entering the science, technology, engineering, and mathematics field disciplines.\textit{Dissertation}

  \bibitem{Madden2021}Madden, L., Carroll, S. Z., \& Schuler, A. K. (2021). Elevating the voices for all learners through shared stories of science learning. \textit{Journal of Science Education for Students with Disabilities, 24(1), 3. https://repository.rit.edu/jsesd/vol24/iss1/3/}

  \bibitem{Pfeifer2020}Pfeifer, M. A., Reiter, E. M., Hendrickson, M., \& Stanton, J. D. (2020). Speaking up: A model of self-advocacy for STEM undergraduates with ADHD and/or specific learning disabilities. \textit{International Journal of STEM Education, 7, 1-21. https://doi.org/10.1186/s40594-020-00233-4}

  \bibitem{Marino2020}Marino, M. T., Vasquez, E., Banerjee, M., Parsons, C. A., Saliba, Y. C., Gallegos, B., \& Koch, A. (2020). Coaching as a means to enhance performance and persistence in undergraduate STEM majors with executive function deficits. \textit{Journal of Higher Education Theory and Practice, 20(5), 94-109. https://doi.org/10.33423/jhetp.v20i5.3040}

  \bibitem{Pfeifer2023}Pfeifer, M. A., Cordero, J. J., \& Stanton, J. D. (2023). What I wish my instructor knew: How active learning influences the classroom experiences and self-advocacy of STEM majors with ADHD and specific learning disabilities. \textit{CBE-Life Sciences Education, 22(1), ar2. https://doi.org/10.1187/cbe.21-12-0329}

  \bibitem{Brown2023}Brown, B. (2023). Describing the experiences of STEM majors with attention deficit hyperactivity disorder and/or specific learning disabilities. \textit{Doctoral dissertation, California State Polytechnic University, Pomona}

  \bibitem{Jordan2014}Jordan, J. A., McGladdery, G., \& Dyer, K. (2014). Dyslexia in higher education: Implications for maths anxiety, statistics anxiety and psychological well-being. \textit{Dyslexia, 20(3), 225-240. https://doi.org/10.1002/dys.1478}

  \bibitem{Jansen2017}Jansen, D., Petry, K., Ceulemans, E., Noens, I., \& Baeyens, D. (2017). Functioning and participation problems of students with ASD in higher education: Which reasonable accommodations are effective?. \textit{European Journal of Special Needs Education, 32(1), 71-88. https://doi.org/10.1080/08856257.2016.1254962}

  \bibitem{Heiman2004}Heiman, T., \& Kariv, D. (2004). Coping experience among students in higher education.\textit{Educational Studies, 30(4), 441-455. https://doi.org/10.1080/0305569042000310354} 

  \bibitem{Taylor2020}Taylor, C. L., Esmaili Zaghi, A., Kaufman, J. C., Reis, S. M., \& Renzulli, J. S. (2020). Divergent thinking and academic performance of students with attention deficit hyperactivity disorder characteristics in engineering. \textit{Journal of Engineering Education, 109(2), 213-229. https://doi.org/10.1002/jee.20310}

  \bibitem{Cai2016}Cai, R. Y., \& Richdale, A. L. (2016). Educational experiences and needs of higher education students with autism spectrum disorder. \textit{Journal of autism and developmental disorders, 46, 31-41. https://doi.org/10.1007/s10803-015-2535-1}
  
  \bibitem{Pfeifer2021}Pfeifer, M. A., Reiter, E. M., Cordero, J. J., \& Stanton, J. D. (2021). Inside and out: Factors that support and hinder the self-advocacy of undergraduates with ADHD and/or specific learning disabilities in STEM. \textit{CBE-Life Sciences Education, 20(2), ar17. https://doi.org/10.1187/cbe.20-06-0107}

  \bibitem{Gin2021}Gin, L. E., Guerrero, F. A., Brownell, S. E., \& Cooper, K. M. (2021). COVID-19 and undergraduates with disabilities: Challenges resulting from the rapid transition to online course delivery for students with disabilities in undergraduate STEM at large-enrollment institutions. \textit{CBE-Life Sciences Education, 20(3), ar36. https://doi.org/10.1187/cbe.21-02-0028}

  \bibitem{Heiman2003}Heiman, T., \& Precel, K. (2003). Students with learning disabilities in higher education: Academic strategies profile. \textit{Journal of learning disabilities, 36(3), 248-258. https://doi.org/10.1177/002221940303600304}

  \bibitem{Graves2011}Graves, L., Asunda, P. A., Plant, S. J., \& Goad, C. (2011). Asynchronous Online Access as an Accommodation on Students with Learning Disabilities and/or Attention-Deficit Hyperactivity Disorders in Postsecondary STEM Courses. \textit{Journal of Postsecondary Education and Disability, 24(4), 317-330. https://eric.ed.gov/?id=EJ966132}

  \bibitem{Mulhall2024}Mulhall, M. (Jul 2024) Students with autism spectrum condition attending higher education: challenges for students and educators. \textit{Dissertation}
  
  \bibitem{Burek2022}Burek, B. L. (2022). Online Research and Comprehension Process and Experience for Postsecondary Students with Learning and Attention Disorders \textit{Doctoral dissertation, University of Toronto (Canada).} 
\end{thebibliography}
\end{document}